\documentclass[aps,prl,amssymb,showpacs,floatfix,preprint,superscriptaddress,nofootinbib]{revtex4-1}
\usepackage{amsmath,amsfonts,graphics,epsfig,amssymb}

\begin{document}
\title{%
IceCube astrophysical neutrinos without a spectral cutoff and ($10^{15}$
-- $10^{17}$)~eV cosmic gamma radiation. }

\author{Oleg Kalashev and Sergey Troitsky}

\affiliation{Institute for Nuclear
Research of the Russian Academy of Sciences, 60th October Anniversary
Prospect 7a, Moscow 117312, Russia}

\email{st@ms2.inr.ac.ru}


\date{October 9, 2014}

\begin{center}
\begin{abstract}
We present a range of unbroken power-law fits to the astrophysical-neutrino
spectrum consistent with the most recent published IceCube data at the
68\% confidence level. Assuming that the neutrinos originate in decays of
pi mesons, we estimate accompanying gamma-ray fluxes for various
distributions of sources, taking propagation effects into account.
We then briefly discuss existing experimental results constraining PeV to
EeV diffuse gamma-ray flux and their systematic uncertainties. Several
scenarios are marginally consistent both with the KASKADE and CASA-MIA
upper limits at ($10^{15}$ -- $10^{16}$)~eV and with the EAS-MSU tentative
detection at $\sim 10^{17}$~eV, given large systematic errors of the
measurements. Future searches for the diffuse gamma-ray background at
sub-PeV to sub-EeV energies just below present upper limits will give a
crucial diagnostic tool for distinguishing between the Galactic and
extragalactic models of the origin of the IceCube events.
\end{abstract}
\end{center}
\maketitle

The observation of an excess of high-energy neutrinos above the
atmospheric background by the IceCube observatory
\cite{IceCube1, IceCube2, IceCubeData} gave a strong boost to
astroparticle
physics (see e.g.\ Ref.~\cite{A-rev} for a review and references). While a
firm conclusion about the astrophysical origin of these events would
require future studies and a confirmation by an independent experiment,
numerous scenarios have been put forward to explain the observation. Not
surprisingly, present low statistics does not allow to single out a unique
explanation of the origin of these events.

Absence of observed events with energies $E \gtrsim 3$~PeV is often
considered as an argument for the presence of a spectral cutoff at these
energies. Indeed, the experimental exposure for electron antineutrinos
peaks around $\sim 6.3$~PeV because of the Glashow resonance, and one
would expect additional events while none is detected (see e.g.\
Ref.~\cite{resonometer} for a detailed discussion). This cutoff would add
further uncertainty to astrophysical explanations because the maximal
energy of 3~PeV is not singled out by any general argument. In this note,
we assume that the neutrino spectrum continues beyond the highest observed
energies and the absence of events at the Glashow resonance is a
statistical fluctuation, not an indication of a cutoff. We will see that
this assumption agrees with the data perfectly.

The general conventional model for production of energetic astrophysical
neutrinos implies their creation in decays of charged pi mesons,
$\pi^{\pm}$, produced in turn in high-energy hadronic or photohadronic
interactions. These $\pi^{\pm}$'s are necessary accompanied by neutral
$\pi^{0}$'s which decay to photons. The energetic photons, therefore, have
to accompany energetic neutrinos, see e.g.\
Refs.~\cite{Gupta-gamma, A-exposure} for
discussions and estimates and
Refs.~\cite{Kohta, Kohta1, Walter} for more detailed model analyses in the
context of the IceCube result. Since the neutrinos propagate freely
through the Universe while the photons may be absorbed, a comparison of
the two fluxes may give important information about the distribution of
sources. In what follows, we will estimate the gamma-ray flux expected in
various scenarios, starting from the IceCube data. Throughout the paper,
we will assume that:
\begin{itemize}
 \item
the neutrino spectrum follows an unbroken power law,
\begin{equation}
\frac{dF}{dE}=N \left(\frac{E}{\rm TeV}   \right)^{-\alpha},
\label{Eq:plaw}
\end{equation}
where the diffuse flux $F$ is measured in cm$^{-2}$s$^{-1}$sr$^{-1}$;
 \item
all neutrinos originate from $\pi^{\pm}$ decays, and the mechanism
producing these $\pi^{\pm}$'s provides for equal amounts of $\pi^{+}$,
$\pi^{-}$ and $\pi^{0}$;
\item
the neutrino mixing is maximal, so, given previous assumptions, the
neutrinos arrive to the observer with $1:1:1$ flavor ratio.
\end{itemize}
Deviations from these assumptions are certainly present but their account
is beyond the precision of the present data.

To proceed, we first need to quantify the astrophysical neutrino flux
consistent with observations. This part of the study may be useful for
other phenomenological considerations, so we discuss it in some detail.

IceCube reported 37 high-energy astrophysical neutrino candidate events at
the background expectation of $\sim 15$. This small statistics is expected
to agree with various descriptions of the spectrum. The original IceCube
paper \cite{IceCubeData} does not present a range of
allowed spectral fits and quotes only two benchmark fits, one with fixed
$\alpha=2.0$ and the best fit with $\alpha=2.3$. However, all necessary
information to obtain the allowed region of
parameters $(N, \alpha)$ at a given confidence level is published, and we
will use it in this study. Namely, the energies of 36 of the observed
events and the energy-dependent number of background events are given in
Ref.~\cite{IceCubeData} while the energy dependence of the exposure is
presented in Ref.~\cite{A-exposure}. There are however two subtle points
in the fitting procedure.

Firstly, the number of events in the sample, and especially in high-energy
bins, is so small that the standard chi-square method would give a biased
result, and the binned Poisson likelyhood does not follow the chi-square
distribution. This is easy to cure, however, with the Monte-Carlo
procedure described e.g.\ in Ref.~\cite{NR}, which is used here.

Secondly, individual neutrino energies cannot be measured and it is only
the deposited energy in the ice which is reported. For cascade events (to
which three highest-energy ones belong), the difference between the two
energies may be neglected, while for track events the deposited energy
gives only a lower limit on the true neutrino energy, which may be several
times higher \cite{IceCubeEnergy}. However, the present IceCube data allow
for a nice trick to overcome this problem, suggested in
Ref.~\cite{A-exposure}. It exploits the fact that, occasionally or not,
there are no observed events with 400~TeV$\lesssim E \lesssim 1$~PeV,
while all three events above 1~PeV are showers. This means that, no matter
shower or track, the energies of all neutrinos except these three do not
exceed 1~PeV. We therefore adopt the approach of Ref.~\cite{A-exposure}
and use only three bins in the analysis: $E<1$~PeV, 1~PeV$<E<$2~PeV and
$E>2$~PeV. We use events with $E>40$~TeV in the fits.

\begin{figure}
\centering
\includegraphics[width=0.95\columnwidth]{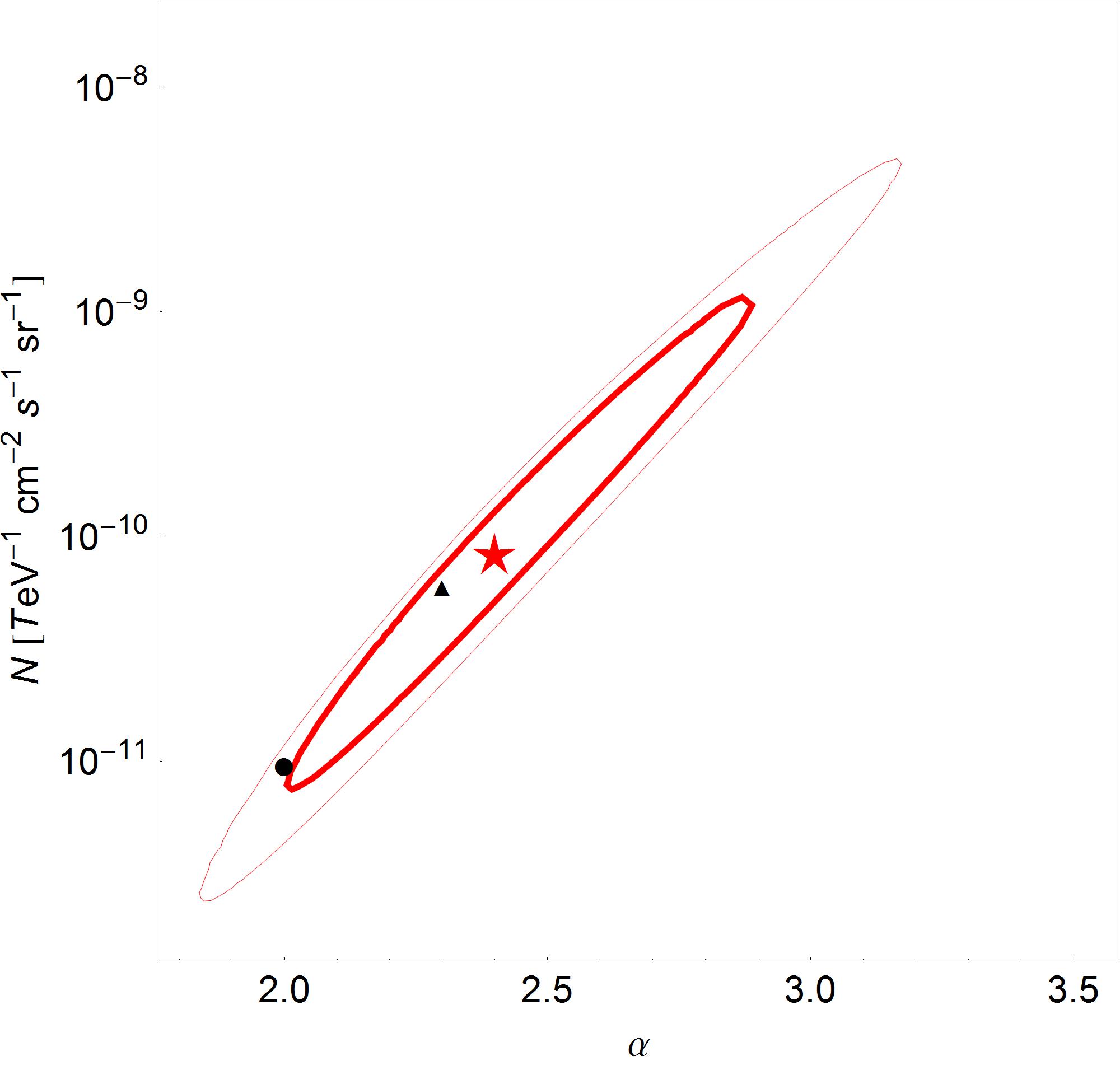}
\caption{
The parameter space (normalization $N$ versus spectral index $\alpha
$) for unbroken power-law fits, Eq.~(\ref{Eq:plaw}), of the astrophysical
neutrino spectrum. The thick and thin contours bound the regions of
parameters consistent with the IceCube data \cite{IceCubeData} at the
68\% and 95\% C.L., respectively. The star denotes our best-fit value,
Eq.~(\ref{Eq:best-fit}). The triangle and the circle denote benchmark
fits of Ref.~\cite{IceCubeData} with $\alpha=2.3$ and 2.0, respectively.}
\label{fig:fit-param}
\end{figure}

\begin{figure}
\centering
\includegraphics[width=0.95\columnwidth]{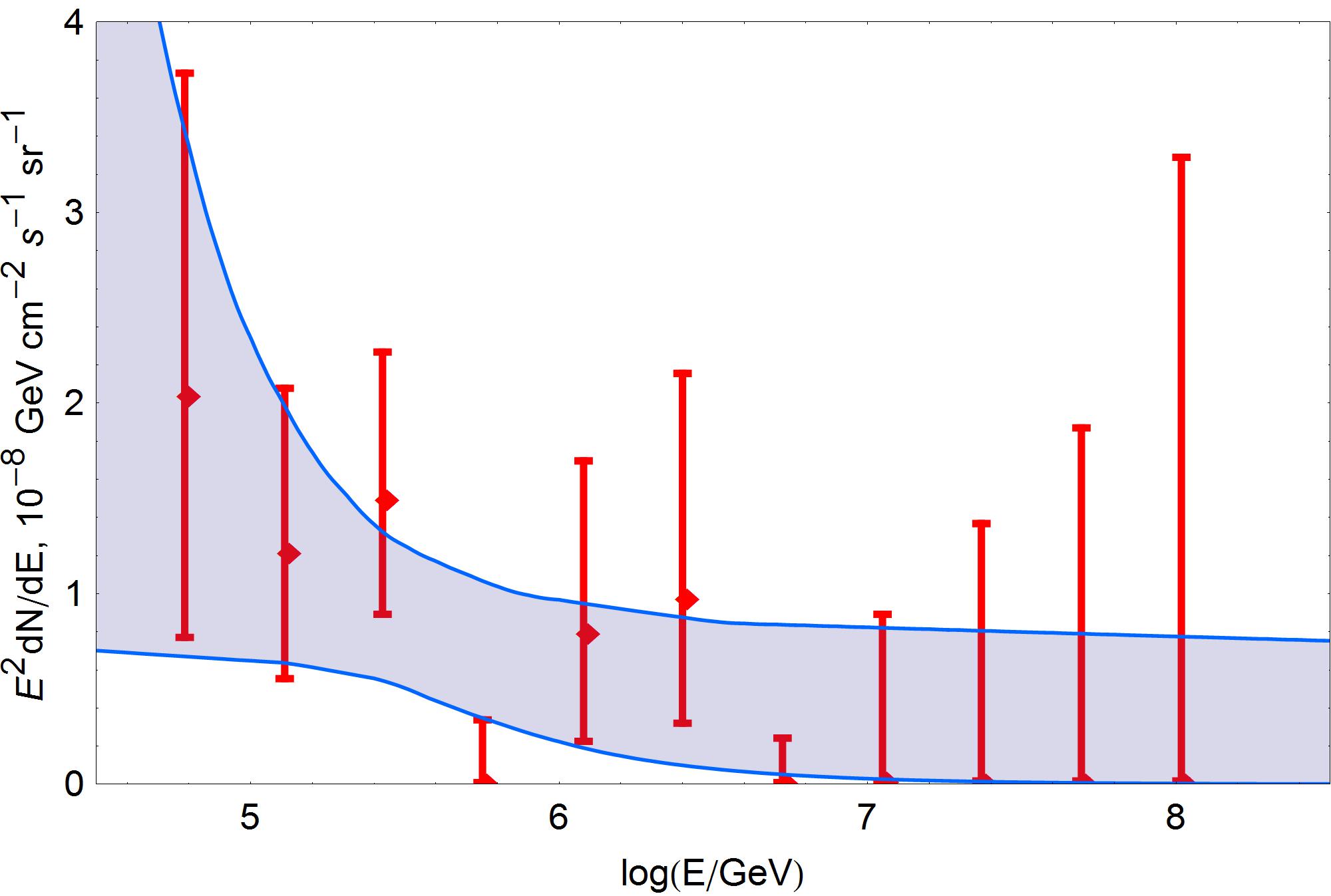}
\caption{
The IceCube astrophysical neutrino spectrum
\cite{IceCubeData} (data points)
together with the range of 68\% C.L.\ allowed power-law fluxes, determined
in Fig.~\ref{fig:fit-param} (shadow).
}
\label{fig:nu-spec}
\end{figure}

The allowed ranges of the spectral fits are presented in
Figs.~\ref{fig:fit-param}, \ref{fig:nu-spec}.
The best-fit values are:
\begin{equation}
\alpha=2.4, ~~~ N=8.7\times 10^{-8}~\mbox{TeV}^{-1} \mbox{cm}^{-2}
\mbox{s}^{-1} \mbox{sr}^{-1} ,
\label{Eq:best-fit}
\end{equation}
only slightly different from the best fit of Ref.~\cite{IceCubeData} (the
difference may originate in the data included in the fit, $E>40$~TeV vs.\
$E>60$~TeV, and in some details of the fitting procedure which is not
described in Ref.~\cite{IceCubeData}). The data are well described by a
power law without any cutoff for a wide range of spectral indices.

Within our assumptions, it is easy to estimate the accompanying flux of
photons from $\pi^{0}$ decays. A simple estimate of the
gamma-ray flux injected by an optically thin source is given by
\cite{A-rev, A-2004}
\[
%{\rm e}^{-\tau(E_{\gamma})}
\left. \frac{dF_{\nu}^{(i)}}{dE_{\nu}}
\right|_{E_{\nu}=E_{\gamma}/2} =
2 \left. \frac{dF_{\gamma}}{dE_{\gamma}}\right|_{E_{\gamma}}
,
\]
where $F_{\nu}^{(i)}$ and $F_\gamma$ are the fluxes of neutrino (per
flavor, that is $1/3$ of the total neutrino flux within our assumptions)
and photons at energies $E_\nu$ and $E_\gamma$, respectively.

On their way from the source to the observer, energetic gamma rays
participate in electromagnetic cascades, driven by the electron-positron
pair production on cosmic background radiations (the relevant
contributions here come from the cosmic microwave, infrared and
ultraviolet backgrounds, depending on the gamma-ray energies) and by the
inverse Compton scattering of electrons and positrons, which produces
secondary energetic photons. Therefore, besides gamma rays from $\pi^{0}$
decays, also electrons and positrons from $\pi^{\pm}$ decays contribute to
the cascade and should be taken into account (their contribution is more
important for lower-energy, $<$PeV, part of the observed spectrum and for
hard spectral indices).
The injected flux $F_{e}$ of electons and positrons at the energy $E_{e}$
is equal to neutrino (per flavour) flux, in the same approximation:
\[
\left. \frac{dF_{e}}{dE_{e}}
\right|_{E_{e}=E_{\nu}} =
 \left. \frac{dF_{\nu}}{dE_{\nu}}\right|_{E_{\nu}}
.
\]

The photon attenuation length
\cite{Lee} is as short as $\sim$10~kpc for
PeV photons, increasing rapidly at both lower and higher energies.
Therefore, mostly Galactic sources may contribute to the gamma-ray flux at
PeV--EeV energies, and the observed spectrum of these gamma rays is very
sensitive to the distribution of sources in the Galaxy and in its
immediate neighbourhood.
%%%%%%%%%%%%%%%%%%%%%%%%%%%%%%%%%%%%%%%%%%%%%%%%%%%%%%%%%%%%%%%%%%%%%%%%%%
In what follows, we use publicly available transport equation based code
written by one of us~\cite{code1, code2} to simulate electron-photon
cascade propagation. In the case of Galactic source distribution for
simplicity we neglect interactions of photons with Galactic infrared and
optical backgrounds. This may lead to at most $5\%$ error in the
$\gamma-$ray flux predictions in the energy range
30~TeV$<E_{\gamma}<$300~TeV only. We also take into account synchrotron
losses of electrons in the $\sim 10^{-6}G$ galactic magnetic field. For the
extragalactic infrared and optical background we use the estimate of
Ref.~\cite{Kneiske:2003tx}.
%we apply Eq.~(\ref{Eq:flux}) to

We use several benchmark source
distributions $n({\bf r})$ in the Unverse. We measure ${\bf r}=(x,y,z)$
from the Galactic Center and account for a non-central position of the Sun
in the Galaxy (assuming that the Sun is 8.5~kpc from the center).

{\bf 1. Stellar distribution}: assume that $n({\bf r})$ follows the
distribution of stars in the Galaxy \cite{astro-ph/0510520}:
\[
n(\rho,z)=\mbox{const} \times \left(n_{\rm d}(\rho,z)+n_{\rm h}(\rho,z)
\right),
\]
\[
n_{\rm h} (\rho,z) =f_{\rm h}  \left(\frac{R_{\rm
Sun}}{\sqrt{\rho^{2}+\left(z/q_{\rm h} \right)^{2}}} \right)^{N_{\rm h}},
\]
\[
n_{\rm d}(\rho,z) =
n_{\rm 1}(\rho,z,L_1,H_{1}) +
f_{\rm f} n_{\rm 1}(\rho,z,L_2,H_{2}),
\]
\[
n_{\rm 1}(\rho,z,L,H) = \exp \left(\frac{R_{\rm Sun}-\rho}{L}
-\frac{|z|+Z_{\rm Sun}}{H} \right),
\]
where the parameter values are $\rho=\sqrt{x^{2}+y^{2}}$, $R_{\rm
Sun}=8.5$~kpc, $Z_{\rm Sun}=2.5$~pc, $f_{\rm f}=0.12$, $f_{\rm h}=0.0051$,
$q_{\rm h}=0.64$, $N_{\rm h}=2.77$, $L_{1}=2.6$~kpc, $L_{2}=3.6$~kpc,
$H_{1}=0.3$~kpc, $H_{2}=0.9$~kpc and  we assume that the distribution
extends up to $r_{\rm max}=15$~kpc.

{\bf 2. Dark-matter distribution}: use the Navarro-Frenck-White (NFW,
Ref.~\cite{NFW}) distribution for $n({\bf r})$, with parameters favoured
by Ref.~\cite{NFW-MW} for the Milky Way:
\[
n(r)
=
\frac{\mbox{const}}{x\left(1+x   \right)^{2}},
\]
where $x=r/R_{s}$, $R_{s}=R_{v}/C_{v}$, the parameter values are
$C_{v}=12$ and $R_{v}=260$~kpc; the integration is extended up to
$r_{\rm max}= R_{v}$.

{\bf 3. Halo hot-gas distribution}: the observed neutrinos may originate
\cite{Kohta, Aha:nu-halo} in interactions of cosmic rays with the hot gas
which probably fills the outer (up to hundreds of Megaparsecs!) halo of
the Galaxy \cite{gas}. For the gas density, we use the Maller--Bullock
\cite{MB} distribution,
\[
n_{\rm gas} = \mbox{const} \times \left(1+
\frac{3.7}{x} \log \left(1+x \right)
-
\frac{3.7}{C_{c}} \log \left(1+C_{c} \right)
\right)^{3/2}, ~~~ r<R_{c},
\]
\[
n_{\rm gas} = \mbox{const}/r^{2}, ~~~R_{c}<r<R_{v},
\]
where $C_{c}=R_{c}/R_{s}$ is the parameter of the gas concentration while
other notations are determined for the NFW distribution.
We have to convolve this distribution with the assumed cosmic-ray density
$n_{\rm CR}({\bf r})$ to obtain the source distribution,
\[
n({\bf r})=n_{\rm gas}({\bf r}) n_{\rm CR}({\bf r}).
\]
In our numerical example, we assume, following Ref.~\cite{Aha:nu-halo},
$n_{\rm CR} \propto 1/r$, and  $C_{c}=1$. Clearly, the both choices are
oversimplified; the real distributions of both the hot gas and cosmic rays
are much more complicated but unfortunately not firmly known. In
particular, both are not expected to be spherically symmetric, and the
photon and neutrino fluxes are expected to be anisotropic in a realistic
case. Deviations from isotropy, which are present also in other scenarios
due to a non-central position of the Sun in the Galaxy, are not discussed
here.

{\bf 4. Extragalactic distribution}: this is used for illustrative
purposes only and assumes $n=$const up to the event horizon. Realistic
source distributions are not continuous and would include the distance to
the nearest source as a parameter, thus reducing the expected gamma-ray
flux even further.

\begin{figure}
\centering \includegraphics[width=0.95\columnwidth]{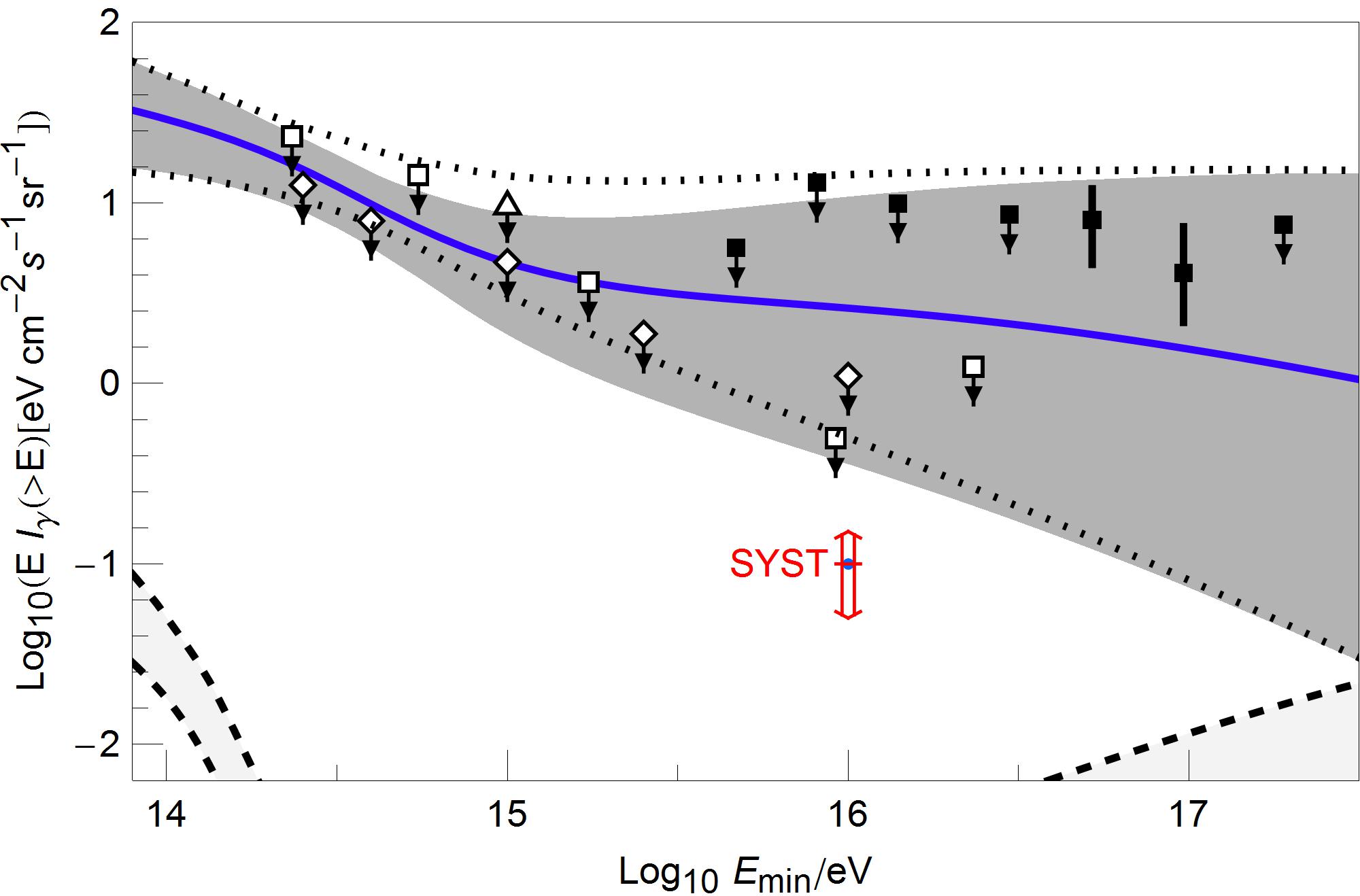}
\caption{
\label{fig:flux}
The diffuse cosmic photon integral flux versus the photon minimal
energy. The dark shadowed region gives the prediction for the source
distribution similar to the hot-gas distribution in the outer halo of the
Galaxy
and for the
68\% C.L.\ range of neutrino spectra. The full line (blue online)
corresponds to the best-fit neutrino spectrum, Eq.~(\ref{Eq:best-fit}),
and the same source model. Predictions for the NFW model are similar to
the hot-gas model, within the plot precision. Dotted lines bound the
range of predictions for the stellar distribution. The light shadowed
regions in the lower part (bound by dashed lines) correspond to a
uniform extragalactic distribution. Experimental constraints are
indicated by symbols:
an open triangle
(EAS-TOP~\cite{EAS-TOP}), open squares (CASA-MIA~\cite{CASA-MIA}),
open diamonds (KASCADE~\cite{KASCADE, MMWG}) and full boxes
(EAS-MSU~\cite{EAS-MSU-flux, EAS-MSU-flux1}). A range of estimated
systematic
errors of the upper limits and measurements is indicated by a double
arrow (red online). }
\end{figure}

The results of the calculations are presented
in Fig.~\ref{fig:flux}.

Several experimental constraints on the gamma-ray flux in the sub-PeV to
sub-EeV energy range are available. They are also presented in
Fig.~\ref{fig:flux} and include upper limits from the
EAS-TOP~\cite{EAS-TOP}, KASCADE~\cite{KASCADE, MMWG} and
CASA-MIA~\cite{CASA-MIA} experiments at energies $\sim (0.1-30)$~PeV as
well as recently published detection claims and upper limits from the
EAS-MSU~\cite{EAS-MSU-flux, EAS-MSU-flux1} experiment at $\sim
(50-500)$~PeV. The main source of systematic errors for all these
results is related to simulations of the background of photon-like
hadronic events, subject to uncertainties of the hadronic interaction
models used. It has been estimated in Ref.~\cite{EAS-MSU-flux}, by
comparison of simulations within various models, as $\pm 50\%$. Though
this kind of an estimate was not presented in earlier studies (in
particular, for the most stringent CASA-MIA limits the
publication~\cite{CASA-MIA} even does not mention which hadronic model was
used for the background estimates), one should expect a similar value of
uncertainty for all experimental results presented in
Fig.~\ref{fig:flux}\footnote{Note that the uncertainty in the energy
estimation, crucial for cosmic-ray studies, is irrelevant here because
the energy of a primary gamma ray is determined in a more or less
model-independent way.}.
These large systematic uncertainties may be responsible for the apparent
tension between CASA-MIA limits at $\sim 10$~PeV and EAS-MSU
detection claims at $>50$~PeV in the context of the scenarios discussed
here, so that a certain hard-spectrum model (the hard spectrum is favored
also by Refs.~\cite{Kohta1, Walter1}) may be consistent with all gamma-ray
constraints. Alternatively, a mixture of Galactic and extragalactic
contributions may result in a better agreement with data.

In any case, our results suggest that PeV to EeV photons represent
a powerful tool to distinguish between models of the origin of IceCube
astrophysical neutrinos and to trace the distribution of their sources.
Galactic models predict gamma-ray fluxes just below, or at the level of,
current observational limits, so searches for primary photons at these
energies are more than motivated. In particular, low-energy extensions of
large cosmic-ray observatories~\cite{TALE, AMIGA} or dedicated
experiments~\cite{HiSCORE} may explore the higher-energy part of the range
very soon.

\begin{acknowledgments}
We are indebted to Grigory Rubtsov for interesting and useful discussions
and to the EAS-MSU group for sharing the results of
Ref.~\cite{EAS-MSU-flux1} prior to publication. This work was supported by
the Russian Science Foundation, grant 14-12-01340. Numerical calculations
have been performed at the computer cluster of the Theoretical Physics
Division of the Institute for Nuclear Research of the Russian Academy of
Sciences.
\end{acknowledgments}

\end{document}